# Single-material 4D-printed shape-morphing structures via spatially patterned strain trapping

S M Asif Iqbal[1,2], Hang Zhang[3], Lin Yang[2,4], Aoyi Luo[2,4], Joseph D. Paulsen[1,2,5], James H. Henderson[2,6]

[1] Department of Physics, Syracuse University, Syracuse, NY, USA
[2] Syracuse BioInspired Institute, Syracuse University, Syracuse, NY, USA
[3] Digital Building Technologies, Institute of Technology in Architecture, ETH Zurich, Zurich, Switzerland
[4] Department of Mechanical and Aerospace Engineering, Syracuse University, Syracuse, NY, USA
[5] Department of Physics, St. Olaf College, Northfield, MN, USA
[6] Department of Biomedical and Chemical Engineering, Syracuse University, Syracuse, NY, USA

**Abstract**
A single-step, single-material 4D printing method is developed for programmable structures featuring spatially patterned strain trapping for one-way actuation. This approach enables fabrication on desktop fused filament fabrication 3D printers through a recently developed shape-memory strain programming method, Programming via Printing (PvP), which eliminates the need for secondary post-fabrication programming. Large (up to 50%) and spatially controlled trapped tensile strain programming is achieved by PvP model design, geometric coding, and printing parameter optimization. While contraction naturally arises from printing-induced trapped strain, expansion is introduced via architected lattice designs with patterned strain — enabling a full range of deformation modes. These capabilities, validated at the unit-cell level, are further integrated into larger proof-of-concept structures to demonstrate scalability and practical implementation. This strategy provides an accessible, low-cost, and easily adoptable additive manufacturing approach for diverse functional-material applications.

## 1. Introduction

Shape Memory Polymers (SMPs) have attracted significant interest for applications in soft robotics, biomedical devices, and aerospace systems due to their large recoverable strain, low density, and tunable mechanical and thermal properties [1]-[9]. SMPs can "memorize" a permanent shape during fabrication through covalent or physical cross-linking. After fabrication, they can be heated, mechanically manipulated into a temporary shape, and cooled to fix (program) that temporary shape via vitrification or crystallization. Subsequent exposure to an external stimulus, such as heat, light, or a magnetic field, triggers recovery to the permanent shape [10]-[15]. SMP-based metamaterials, in which lattice properties are influenced by the microstructural designs, have been studied, but much of their potential remains unrealized due to the need for a post-fabrication programming step, which has limited the programming of both intricate geometries, which can be difficult to manipulate into the desired temporary shape, and of precise, complex shape changes, such as those involving multi-axial manipulation [10], [13].

To address limitations associated with traditional post-fabrication programming, we and others [16]-[21] have recently demonstrated single-step, single-material fabrication and programming of SMPs by fused filament fabrication (FFF). FFF is a popular extrusion-based additive manufacturing (AM) method for the 3D printing of polymers, in which the polymer is melted and deposited as a fiber, layer-by-layer, on a print bed [22]. In our single-step programming approach, which we refer to as Programming via Printing (PvP), the heated polymer fiber is stretched during deposition, and molecular-level tensile strain is fixed through vitrification as the fiber cools on the print bed, thereby programming shape memory at the fiber level in a single-step, single-material printing and programming process [16], [17]. The magnitude of the trapped strain can be tuned by the FFF parameters, such as nozzle temperature and multiplier (which controls the flow of extruded materials). Importantly, PvP has been developed and implemented using inexpensive, off-the-shelf, hobbyist printers and commercially available SMPs, making the approach an easily adoptable, low-cost, and accessible one.

While these advances represent a significant step forward, key limitations exist. First, the trapped strains achieved to date by PvP and related approaches are generally low when compared to those achieved during traditional programming. For example, when calculated using a directly comparable method (**Equation (1)**) based on strains recovered from the printed shape (for PvP) or temporary shape (for traditional programming), previously reported PvP trapped strains have reached 20–30% [5], [18], [21], [23], while traditional programming has demonstrated trapped strains up to 75% [24], [25], [26]. Only one printing study has reported substantially higher strain (63.1%), but this required a custom-synthesized poly(ether urethane) (PEU) SMP and unusually low nozzle temperature [23]. Achieving high-magnitude trapped strains with commercial SMPs is essential if PvP and similar single-step approaches are to broadly enable applications requiring large deformations or complex shape changes. Second, although strain magnitude can be tuned via FFF process parameters, fiber-level spatial control of strain within a *single structure*—critical for complex shape changes—has not been demonstrated. Finally, PvP and similar methods currently enable only tensile strain trapping, limiting shape changes to contraction-driven behaviors such as uniaxial shrinkage or simple bending. The ability to achieve expansion, particularly multiaxial expansion, remains an unmet need. Addressing these limitations could make broadly available an inexpensive but powerful AM approach for fabricating and programming intricate geometries and complex shape changes in metamaterials.

This work addresses these three challenges by advancing PvP to (i) achieve higher trapped strains (~50%) using commercial SMPs, (ii) enable facile spatial control of fiber-level strain within a single structure, and (iii) realize multiaxial expansion through geometric design strategies. All advances use an inexpensive, single-material, off-the-shelf hobbyist printer and commercially available SMP. Here we report a method for implementing PvP, based on a unit cell that couples contraction of single struts to expansion of the overall structure along one or more axes. The advances are achieved through systematic exploration and optimization of printing parameters, modification of geometric code (G-code), and modeling-based design of geometrical mechanisms that translate contractile tensile strains into expansion. Unit cell concepts are validated and then scaled to proof-of-concept structures, demonstrating the potential of the approach for low-cost, accessible fabrication of complex, programmable architectures.

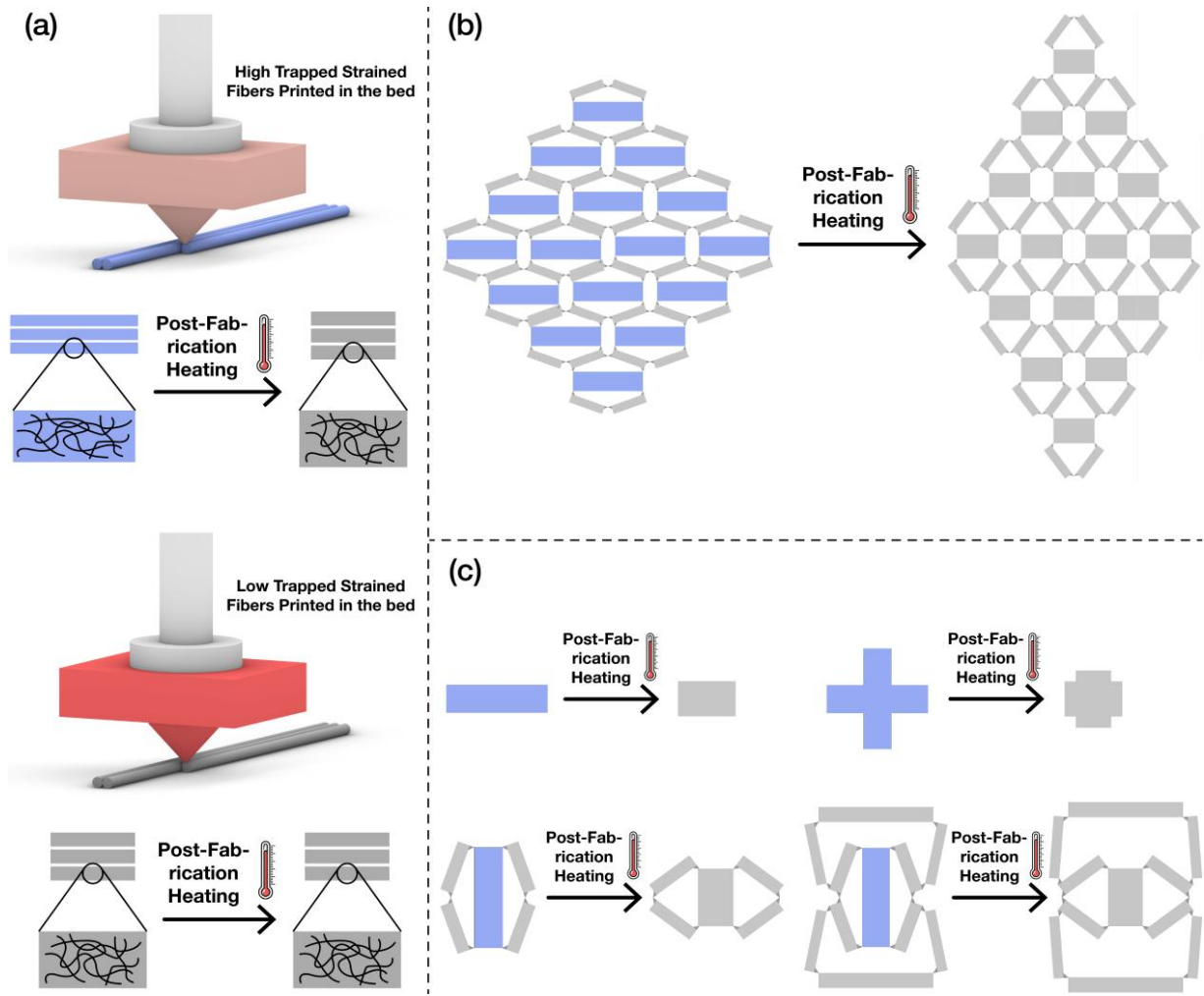

**Figure 1**

(a) Fabrication of active and passive elements with FFF and PvP by controlling the trapped strain (b) Demonstration of large-scale uniaxial expansion through contracting elements in a lattice configuration (c) Conceptually, this work demonstrates the use of smart materials to achieve uniaxial and biaxial expansion through contraction. While uniaxial and biaxial contractions can be achieved directly from the trapped strain, multiaxial expansion requires the combination of fabrication strategy and geometry.

## 2. Results and Discussion

### 2.1. Effect of Printing Parameters and Dimensions on Trapped Strain

Our prior work and that of others using extrusion-based approaches like PvP have shown that lowering nozzle temperature increases strain trapping [17], [18], [19], [23], [27], [28]. In addition, while some reports have found that trapped strain increases with higher printing speed [22], [29], [30], a few other research groups have observed constant trapped strain above a printing speed threshold [18], [31]. A limitation of these prior reports was that the effect of printing speed on trapped strain and bending was explored for a single nozzle temperature. Here, we systematically investigate how nozzle temperature influences the relationship between printing speed and trapped strain.

To address this gap and more comprehensively maximize strain trapping, we systematically investigated the influence of nozzle temperature (195 to 235°C, 20°C increments) and print speed (10 to 60 mm/s, 10 mm/s increments) on strain trapping. At the extremes we explored, we identified practical constraints that highlight the need to balance strain trapping with structural integrity and fidelity. Attempts to print at 190°C were unsuccessful due to poor print quality and reproducibility, despite the polymer's melting onset near 160°C and near-complete melting at 180°C [32]. Narrow SMP strips (20 mm × 1.2 mm × 1.2 mm) were 3D printed from SMP MM3520 filament, each strip comprising six layers in total (each layer 20 mm × 0.2 mm × 0.2 mm). During printing, strain was trapped within the polymer chains as they were stretched and then cooled at room temperature. Narrow strips were chosen to understand and characterize the strain trapping behavior of the narrow beam elements quickly. Post-printing, samples were fully recovered in a hot water bath ($T_g$+35°C = 70°C for 30 min). Trapped strain was quantified by measuring samples immediately after printing and then after recovery.

$$Trapped\ Strain = \frac{L_f - L_i}{L_i} \times 100\% \qquad (1)$$

where $L_f$ = final length after recovery, $L_i$ = initial length after fabrication.

The experimental results indicate that the trapped strain decreases with increasing nozzle temperature and decreasing printing speed (**Figure 2(a)**). The highest trapped strain of 48.7% was observed at the lowest printing temperature and highest printing speed tested (195°C and 60 mm/s), whereas the lowest trapped strain was observed at the highest printing temperature and lowest printing speed tested (235°C and 10 mm/s). Notably, nozzle temperature had a more pronounced effect on strain trapping than printing speed, which exhibited a gradual increase in trapped strain with increasing printing speed (**Figure S 1**, supplementary materials). At 195°C and 235°C, the effect of printing speed on trapped strain was not statistically significant at speeds ≥30 mm/s ($p > 0.05$), whereas at 215°C, an increasing trend was observed even above 40 mm/s. These results demonstrate that the influence of printing speed on trapped strain is temperature-dependent, with lower and higher nozzle temperatures reaching stable regimes earlier than the intermediate temperature.

Although our print design was intended to primarily trap tensile strains only, in some cases we observed moderate bending of structures upon recovery. Bending increased with increasing printing speeds at a given nozzle temperature (**Figure 2(b)**). Bending occurred upward relative to its original position on the build plate, with the convex side corresponding to the topmost printed layer (**Figure S 2**, supplementary materials). The bending direction might be explained by the fact that all the lower layers except the topmost layer get reheated during the FFF process, introducing trapped strain mismatch between the layers [22], [33]. Samples printed at lower printing speeds (10 and 20 mm/s) showed negligible (at 195°C and 235°C) or inconsistently (at 215°C) bending and were excluded from curvature analysis. The radius of curvature did not directly correlate with nozzle temperature.

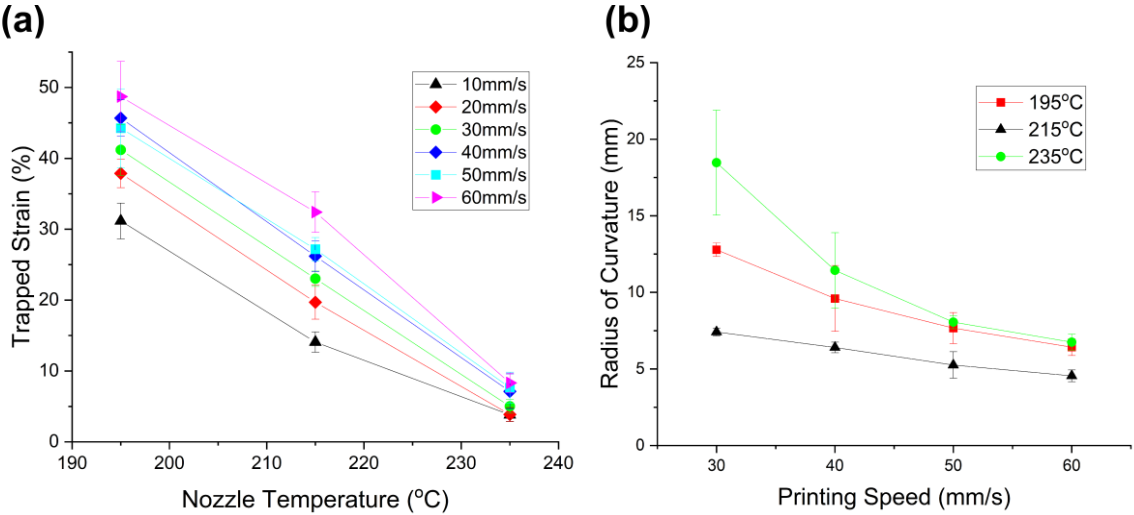

**Figure 2**

(a) The trapped strain for six different printing speeds for three different nozzle temperatures observed in SMP narrow samples. At higher nozzle temperature, the difference in the trapped strain decreased. (b) Radius of curvature as a measure of the extent of bending for different printing speeds and nozzle temperatures. Interestingly, less bending was observed when the nozzle temperature was in the intermediate regime (215°C).

While controllable self-bending is advantageous for many potential applications [21], [31], [34], [35], [36], our objective was to achieve uniaxial shrinkage without bending to enable linear shape-morphing transitions for architected designs. To reduce bending, we modified the strip dimensions to 20 mm × 5.5 mm × 1.2 mm and gradually increased the thickness of the sample up to 3.2 mm (printed at a constant speed, 30 mm/s), reasoning that additional layers would constrain bending as a larger percentage of the layers are away from the substrate effect [29]. The trapped strain again increased with decreasing nozzle temperature, achieving a maximum trapped strain of nearly 60% for beams measuring 20 mm × 5.5 mm × 1.2 mm (at 195°C, the lowest nozzle temperature, **Figure 3(a))**, however, noticeable bending was observed. This trapped strain is among the highest achieved for PvP and related approaches [5], [22], [23]. Increasing thickness resulted in a decrease in trapped strain, and it was reduced to 50% at 195°C (**Figure 3(a)**). However, negligible bending was observed compared to the other thickness groups (1.2 mm and 2.2 mm, **Figure 3(c)**), confirming the feasibility of achieving purely contractile behavior. We also varied the print speed for different nozzle temperatures (**Figure 3(b)**), and a similar trend emerged for other printing speeds (10 and 50 mm/s), reaching up to 55%. Again, no bending was observed across samples, confirming the feasibility of achieving purely contractile behavior. Manufacturer guidelines for TPU-based 4D printing recommend 10–30 mm/s, and we observed that higher speeds compromised layer adhesion and extrusion consistency. Overall, we were able to achieve up to approximately 50% trapped strain while maintaining good print quality and minimal self-bending.

We suspect that due to the substrate effect (here, the print bed) the narrow strips showed comparatively less trapped strain with moderate to large amount of bending. If this issue is managed appropriately, one might be able to get a thinner strip with more trapped strain. However, this is out of scope of this paper and will be future work. Overall, these results provide a set of guidelines for optimized printing and geometric parameters to achieve linearly contractile beams with maximum possible trapped strain with consistent print quality. Next, we turn to using these struts to fabricate shape-shifting structures with PvP.

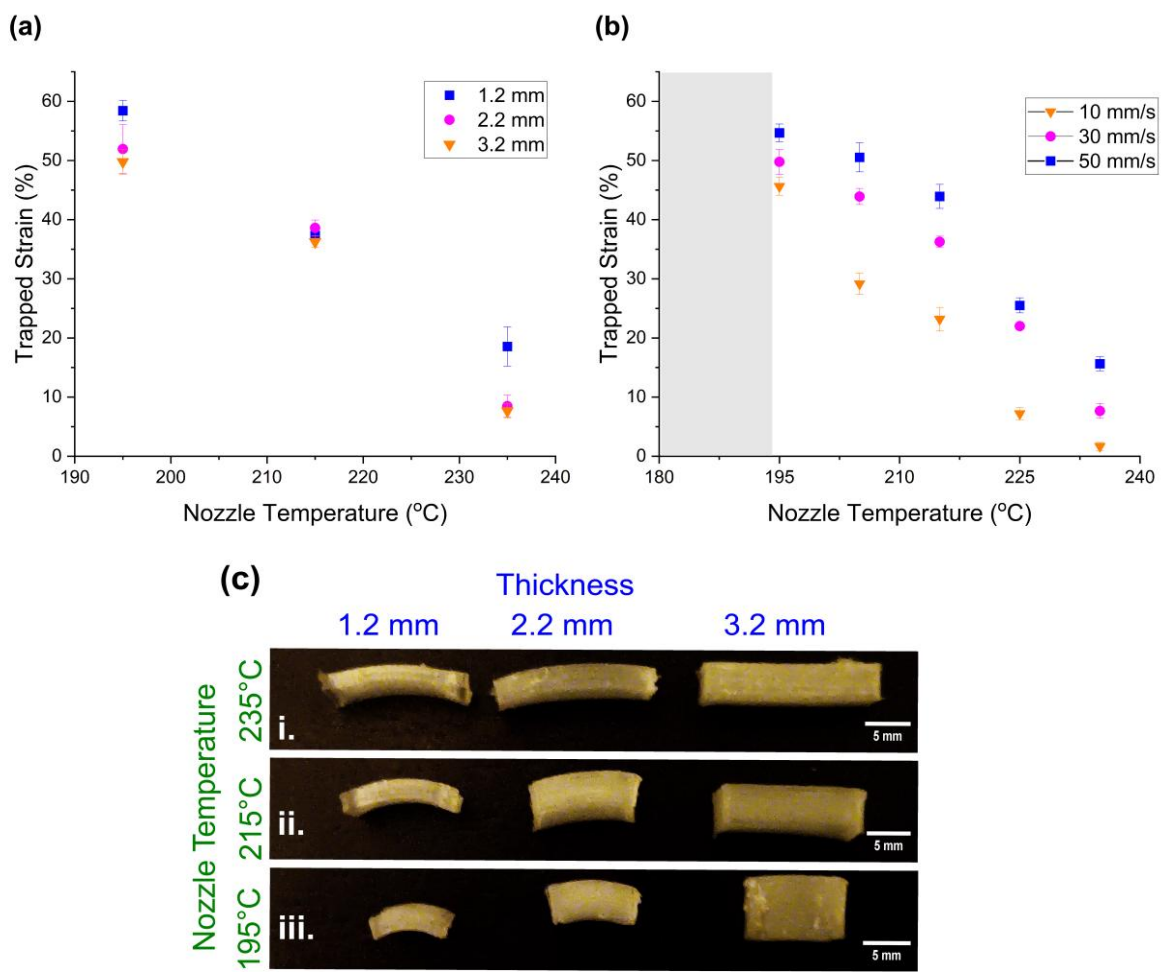

**Figure 3**

(a) Trapped strain when increasing the thickness of an SMP rectangular bar of dimensions 20 mm × 5.5 mm from 1.2 mm to 3.2 mm. The increase in layer number or thickness to 3.2 mm resulted in lower trapped strain but no observed bending. (b) The print speed was then varied for different nozzle temperatures for 3.2 mm thick samples. The results were consistent, showing slight variation in trapped strain but no bending. (c) Post-recovery images of SMP rectangular samples with different thicknesses, when the layer numbers or thickness decreased, there was an increase in trapped strain; they showed small (for 2.2 mm) to moderate (1.2 mm) bending as shown in the optical images. The quantitative data has not been shown. Here, the printing speed was constant for all the samples (30 mm/s).

## 2.2. Uniaxial Expansion via Linear Shape-Morphing Unit Cells

### 2.2.1. Analytical Design and Geometrical Modeling

PvP can print elements that contract after actuation due to trapped tensile strain but cannot directly print elements that extend or expand. To make this printing technique a versatile actuation platform capable of both contraction and extension, extension is achieved through an architected structure design that converts element-level contraction to unit cell-level expansion. Here we show how uniaxial contraction of some beams within a larger architected cell can lead to the effective expansion between specific material points in that structure. We explore this concept by developing an idealized geometric model to predict unit cell behavior, then validate this model experimentally in subsequent sections. To provide a simple yet instructive insight, in the first version of the model, two key assumptions are made: (1) the beam elements are treated as rigid, straight lines of negligible width, (2) their endpoints are connected by ideal rotational joints (**Figure 4(a)**). We assume only the center beam has a trapped strain of $\varepsilon$ and the other beams do not change their dimensions. As shown in **Figure 4(a)**, the initial distance between the top and bottom points of the structure is $d_0$, and the distance after actuation is d (**Equation (2) and (3)**).

$$d_0 = \sqrt{4(L_1)^2 - (L_0)^2}, \tag{2}$$

$$d = \sqrt{4(L_1)^2 - [L_0(1-\varepsilon)]^2}, \tag{3}$$

Here, $L_0$ is the length of the center active element, and $L_1$ is the length of the outer passive element. **Figure 4(c)** and **(d)** show the modeling results for nondimensionalized expansion displacement (($d - d_0$)/$L_0$) and expansion strain (($d - d_0$)/$d_0$) as a function of $L_1/L_0$ for various trapped strains $\varepsilon$. It can be seen that both ($d - d_0$)/$L_0$ and ($d - d_0$)/$d_0$ increase as $L_1/L_0$ decreases and $\varepsilon$ increases. This trend is further supported by evaluating the partial derivatives of ($d - d_0$)/$L_0$ and ($d - d_0$)/$d_0$ with respect to $L_1/L_0$ and $\varepsilon$. It can also be shown that the partial derivative for $\varepsilon$ is always positive (**Equation (S5) and (S8)**), while the partial derivative for $L_1/L_0$ is always negative (**Equation (S6) and (S9)**). As a result, maximum expansion displacement and strain are achieved when $L_1/L_0=0.5$ (since $L_1/L_0$ cannot be smaller than 0.5 due to the geometric constraint of a triangle) for a given $\varepsilon$, and a higher $\varepsilon$ is beneficial to achieve higher expansion displacement and strain.

To evaluate whether the triangular configuration yields the maximum expansion displacement and strain for a given trapped strain $\varepsilon$, we extended the geometric model to a structure composed of two connected trapezoids (discussed in details in the Supplementary Information), as shown in **Figure S 4**. The analytical framework developed for the triangular case is adapted accordingly, as **Equations (S21)-(S25)** show, ($d - d_0$) achieve maximum when the length of the top base of the trapezoid approaches 0, which forms a triangle. This shows that the triangle design gives the maximum expansion when compared to a trapezoid.

The effect of beam width is then incorporated into the geometric model for a more comprehensive understanding, as illustrated in **Figure 4(b)**. In this revised model, the beams are no longer idealized as lines but are represented with finite thickness, and they are connected at their outer edge points, which remain free to rotate.

$$d_0 = d_{0\_ori} + H_0 + \frac{L_0 H_1}{2L_1}, \tag{4}$$

$$d = d_{ori} + H_0(1-\varepsilon)^{-\nu} + \frac{L_0 H_1(1-\varepsilon)}{2L_1}, \tag{5}$$

The initial vertical distance $d_0$ (**Equation (4)**) is now given by the original distance predicted by the simplified model (neglecting beam width), denoted as $d_{0\_ori}$, plus an additional term accounting for the geometric contribution of beam width. Specifically, $H_0$ and $H_1$ represent the widths of the central active beam and the surrounding passive beams, respectively. Similarly, $d = d_{ori}$ (distance predicted by the model neglecting the beam width, plus the additional terms from Poisson's effect (v) and the geometric influence of beam width (**Equation (4) and (5)**).

The effect of beam width is shown in **Figure 4(e)-(h)**, which plots the nondimensionalized expansion displacement ($(d - d_0)/L_0$) and expansion strain ($(d - d_0)/d_0$) as a function of $H_0/L_0$ and $H_1/L_0$. The expansion displacement $(d - d_0)/L_0$ monotonically increases with $H_0/L_0$, as evidenced by its consistently positive partial derivative of $(d - d_0)/L_0$ with respect to $H_0/L_0$ (**Equation (S14)**) While **Figure 4(e)** suggests that $(d - d_0)/L_0$ increases as $L_1/L_0$ decreases, the partial derivative $\partial[((d - d_0)/L_0]/\partial(L_1/L_0)$ can be either positive or negative (**Equation (S16)**), indicating that the relationship is parameter-dependent. Similarly, the expansion strain $(d - d_0)/d_0$ exhibits variable behavior with respect to $H_0/L_0$ (**Figure 4(f)**) with $\partial[(d - d_0)/d_0]/\partial(H_0/L_0)$ being either positive or negative depending on the parameter regime (**Equation (S18)**). The partial derivative $\partial[(d - d_0)/d_0]/\partial(L_1/L_0)$ also shows sign variability (**Equation (S20)**), confirming the indeterminate trend. In contrast, $(d - d_0)/d_0$ decreases monotonically with increasing $H_1/L_0$, as their partial derivatives with respect to $H_1/L_0$ are consistently negative (**Equation (S19)**).

Note that the $((d - d_0)/L_0)$ and $((d - d_0)/d_0)$ for the sole active beam due to Poisson effect during shrinkage is also plotted as a line labeled "$H_0$ only". Except at small values of $L_1/L_0$ and $H_0/L_0$, both $((d - d_0)/L_0)$ and $((d - d_0)/d_0)$ exhibit greater changes than those attributable to Poisson's effect alone.

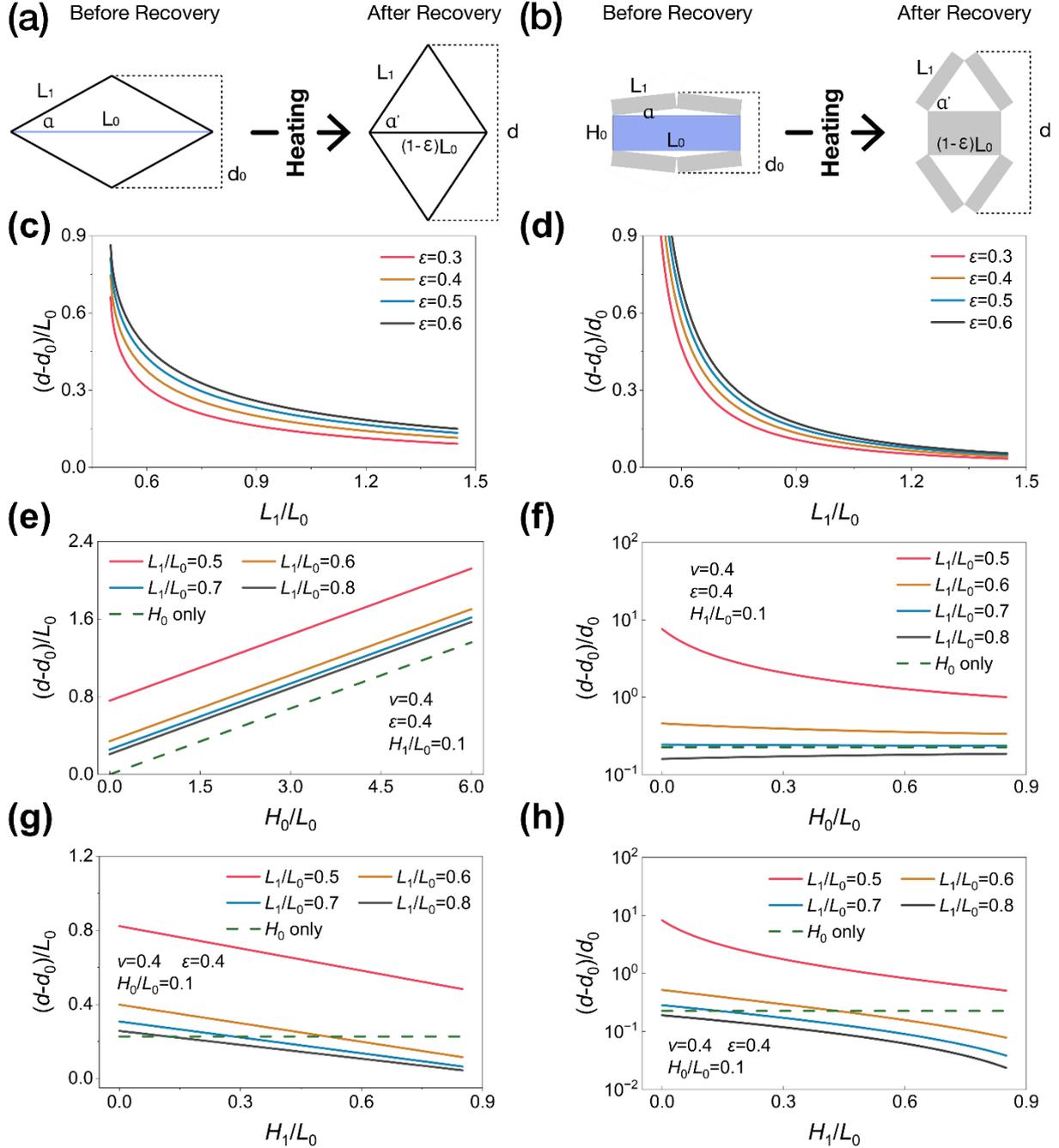

**Figure 4**

(a, b) Sketches of the concept of unit cell; (a) lines representing active (of length $L_0$) and passive elements (of length $L_1/2$), (b) lines are replaced with beams. (c-h) Computational results based on the model depicted in (a) and (b)

### *2.2.2. Fabrication and Demonstration of Uniaxial Expansion*

Based on design and geometric modeling, a PvP-based 4D printing strategy was devised to fabricate the diamond-shaped unit cell. Each unit cell comprises a central active contractile beam ($L_0$, 20 mm × 5.5 mm × 3.2 mm) and four surrounding passive non-contractile beams (e.g., $L_1$, length varied with model, provided in SI, width ~ 2.2 mm and thickness ~ 3.2 mm, **Figure *5*(a) and (b))**, connected by flexible triangular joints (1.6 mm thick) at the lower half of the contractile and non-contractile beams. The triangular-shaped joints were chosen as the connecting nodes since they minimize contact area, allowing the structure to more closely approximate a low-friction hinge. The active-beam dimensions were selected based on **Section 2.1** results, which demonstrated negligible bending at moderate printing speed (30 mm/s) for a wide range of nozzle temperatures. To maximize trapped strain, the active beam was printed at 195°C, while passive beams and triangular joints were printed at 240°C to minimize trapped strain. The entire structure was fabricated in a single print (details in Experimental Section).

Six models (N=3 per group) were fabricated with varying passive beam lengths to assess geometric effects on expansion. The smallest angle, α (≈ 4°), produced nearly parallel active and passive beams, while subsequent models increased α and passive beam length incrementally, as can be seen in representative Computer-Aided Design (CAD) models and printed samples before and after recovery (**Figure *5*(a) and (b)**). Expansion was quantified by measuring the change in height ($d_0$ to d) between the outermost points of the unit cell before and after actuation (**Figure *5*(a) and (b)**), and the percentage of expansion was calculated. The passive beam remained nearly constant ($L_1 \approx L_1'$) during actuation.

The largest expansion (≈75.7%) was achieved for α ≈ 4° ($L_1 \approx L_0$). Increasing α or $L_1/L_0$ resulted in a monotonic decrease in expansion, reaching 27.3% at the largest angle tested, which was lower than the transverse expansion expected from the Poisson effect alone (**Figure S 3**, supplementary materials). Trapped strain in active beams was ≈43%, slightly lower (≈5% less) than the contraction in a free state for 30 mm/s printed samples (Section 2.1). This decreasing trend confirms that longer passive beams reduce vertical displacement for a given active strain.

Two mechanisms drove bulk expansion of the structures: (i) lateral expansion of the active beam due to the Poisson effect, and (ii) outward rotation of passive beams enabled by flexible hinges as the active beam contracted, consequently increasing the overall structural height. While the absolute height of the unit cell increased with α, the percentage expansion decreased because the passive beams contributed negligible axial contraction or lateral expansion. This geometric trade-off highlights the importance of hinge placement and beam proportions in tuning actuation performance.

Finally, the model results were then compared with the experiments. Different from the geometry in **Figure 4(b)** where the beams are connected at their edge points that are free to rotate, the designed geometry for experimental fabrication (**Figure 5(a) and (b)**) has triangular hinges. So the exact geometry, including the triangular hinges in **Figure 5(a)**, is used in the model, assuming beams together with the triangular hinges are rigid, and rotate with respect to the points that the triangular hinges contact. An animation of the deformation of the model is shown in the **Video S1** and the corresponding $d_0$ and d are shown in **Figure 5(c)** and **(d)**. Good agreement is observed, which shows the validity of the model, and that we can consider the behavior as rigid parts connected at rotational points.

This section demonstrates a single-material, single-step printed unit cell capable of large, symmetric uniaxial expansion through programmed spatially controlled trapped strain and geometric amplification. Nozzle temperature control was critical for embedding differential strain in active and passive elements. Although hinges were asymmetrically positioned for fabrication convenience, future designs could leverage mid-air printing or alternative hinge configurations to further optimize performance.

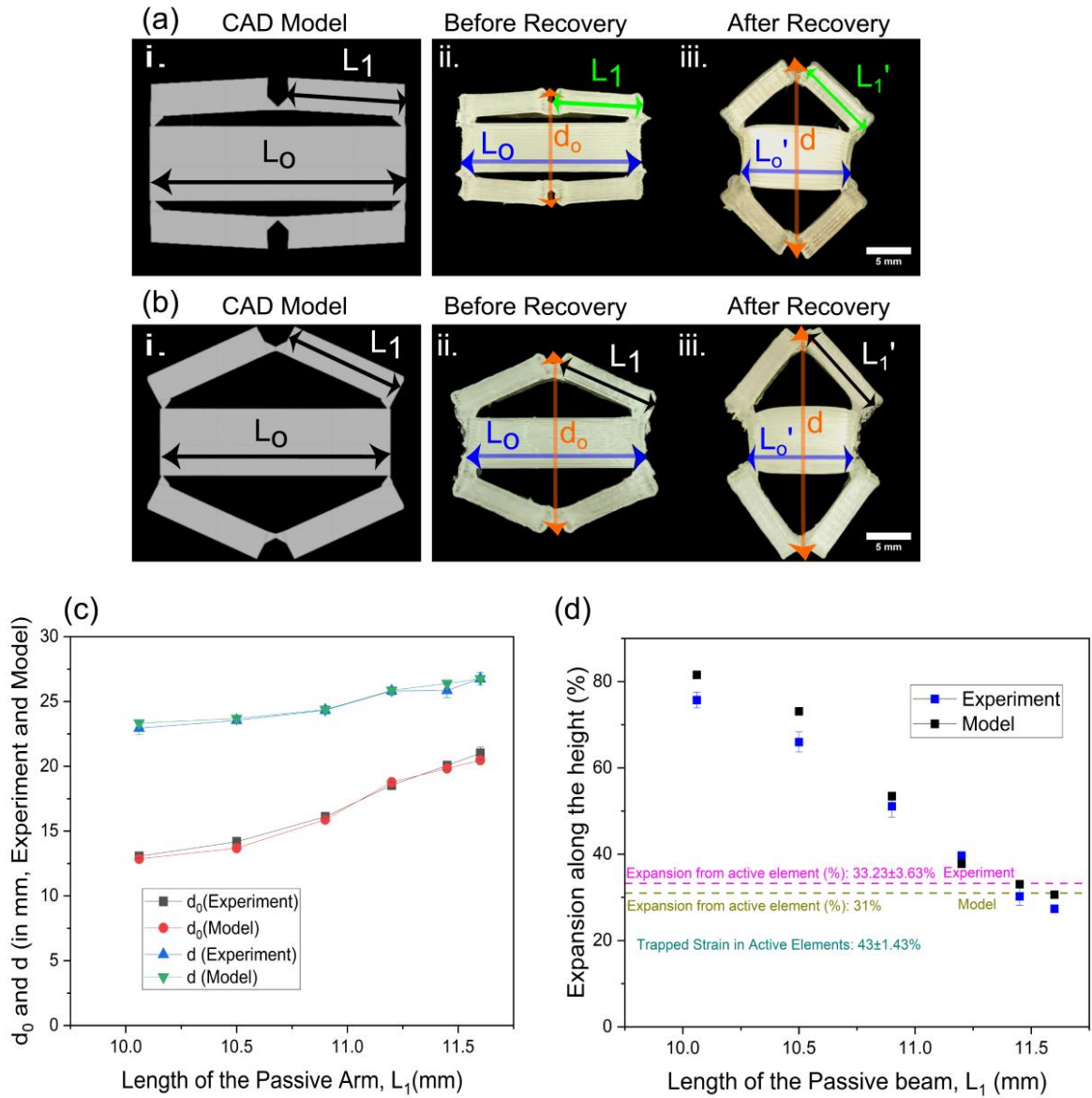

**Figure 5**

(a, b) CAD design of the unit cells and optical images of samples (before and after recovery) from two distinct representative models (where α was different) (c) Comparison of geometrical modeling and experimental values for $d_0$ and d. (d) Experimental results showing expansion for different trapped strains, expansion from the active element was also computed for both the model and the experiment.

### 2.3. Utilizing Trapped Strain for the Demonstration of Biaxial Expansion

#### 2.3.1. Design Strategy and Geometrical Modeling

Building on these results for the uniaxial expansion strategy, we next targeted biaxial expansion of printed structures. To understand the behavior of the designed cell, we adopted the same approach that assumes the structure consists of rigid beams connected with free rotational hinges. The structure is the combination of the diamond-shaped structure studied in **Figure 5** and a reentrant structure to create a biaxial mechanism as shown in **Figure 6(a)**. In this structure, the contraction of the active element determines the expansion of the diamond-shaped structure, which is the distance between the center points of the reentrant structure. This distance then determines the expansion of the auxetic-like structure in the x-direction. Since we already investigated the relationship between the active element contraction and the expansion of the diamond-shaped structure, now we only need to investigate the relationship between the distance between the center points of the reentrant structure (which is just the expansion of the diamond-shaped structure) and the expansion of the auxetic structure in the x-direction. Consider the auxetic structure width w and its relative change Δw.

As can be seen (**Equation (S28)**), Δw is independent of the beam width of the outer reentrant structure ($w_2$ and $w_3$, width of the beams corresponding to $L_2$ and $L_3$). Thus, we use the line model where the reentrant structures are considered as rigid lines connected with ends that can freely rotate (i.e., $w_2 = w_3 = 0$), and **Figure 6(d)** plots $w/L_2$ as a function of $(L_3-d)/L_2$, which is an even function with maximum value at $(L_3-d)/L_2=0$. $\Delta w/L_2$ can be obtained by calculating the y value difference of any two points on the curve. From **Equation (S27) and (S28),** and the curve, it can be seen that for a given $d_0$, the highest expansion is obtained when $L_3=d$, where the reentrant structure becomes a rectangle as shown in **Figure 6(b)**. When $L_3=d$ is satisfied, the larger the $|L_3-d_0|$, the higher the expansion ($\Delta w/L_2$).

### 2.3.2. Fabrication and Experimental Observation

Based on the design discussed in the previous section, an auxetic frame was fabricated around the previously described uniaxially expanding triangular unit cell. Similar to the analytical model, with the joints triangular in shape as described for the unit cell structure, in Section 2.2.2, the auxetic-like structure was fabricated to transmit expansion from the unit cell to the surrounding non-contractile arms, labeled as $L_2$ and $L_3$ (**Figure 6(b)**) and drive motion in the non-contractile arms, enabling expansion in both the x- and y-direction (**Figure 6(a)**). The printed structure consisted of 16 layers (total height: 3.2 mm), fabricated at a printing speed of 30 mm/s. The print direction was aligned with the beam length to maximize shrinkage during recovery. The active element was printed at 195°C, while all other components were printed at 240°C. Non-contractile elements were printed with fibers oriented perpendicular to the long axis to minimize shrinkage during heating. Flexible triangular hinges (1.6 mm thick) connected to the lower half of the elements. For demonstration, the unit cell with the smallest α was selected. Again, all the parts of the structure were printed simultaneously.

Upon heating above the $T_g$, the shrinking beam actuated, transforming the internal angle β from 130° to 180° (**Figure 6(b)**). This transformation resulted in biaxial expansion, particularly along the x-axis. Before recovery, $L_2$ ≈ 12.5 mm and w ≈ 12.25 mm; after recovery, w ≈ 13.7 mm, corresponding to 11.83% expansion along the x-axis and approximately 80% expansion along the y-axis, as the K = (d − $L_3$)/2 → 0 (**Figure 6(c)**). Geometrical analysis (**Figure 6(d)**) supports this behavior: the width is given by $w = 2\sqrt{L_2^2 - K^2}$. When K =0, w = 2$L_2$, representing maximum x-axis expansion. The model predicted w = 14.7 mm, which slightly exceeded the experimental value, likely due to minor fabrication tolerances or hinge compliance.

For better comparison, the exact design used in printing was then employed in modeling, which includes the triangular hinges. The model assumes beams together with triangular hinges are rigid and rotate with respect to the points that the triangular joints contact. Animation of the deformation of the model is shown in **Video S2**.

Collectively, these results demonstrate that biaxial expansion can be achieved using a single contractile beam. A promising design variation involves programming trapped strain in the arm $L_3$, potentially increasing x-axis expansion. Trapping strain in all arms could further enhance expansion, though this would require careful parametric design and process validation of PvP to avoid unintended bending or unpredictable tension distributions.

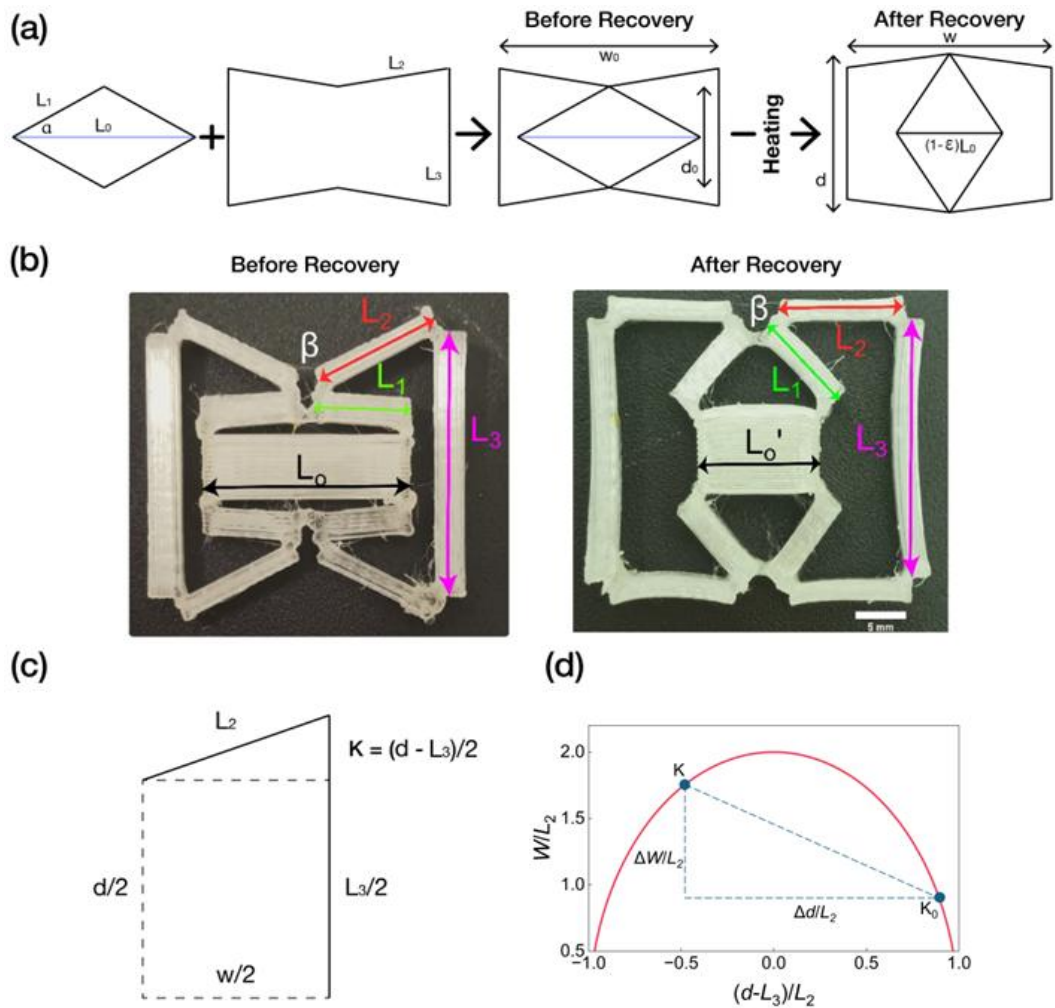

**Figure 6**

(a, b) Before and after recovery images of the samples where the unit cells were inserted within an auxetic configuration. (c, d) Geometric model of the configuration.

## 2.4. Demonstration of Multicellular Structure

Previously, we defined a unit cell as the combination of an active central element and two pairs of passive beams. In this section, to demonstrate scalable, bulk uniaxial expansion, we fabricated a multicellular structure with 9 horizontal active elements (**Figure 7(a)**). In this design, the triangles are open, with the passive beams only connected to the active beam for design simplicity. All other printing conditions and fabrication methods were consistent with those described in earlier sections. This multicellular structure can also be described as a lattice, where a weighted combination of the central contracting active beam and two rotating passive beams (**Figure 7(a) and (b)**, indicated by differently colored shaded regions) was considered.

The active central beam contributed ~ 33% expansion (**Figure *5(d)***) due to the Poisson effect, while the rotating hinges provided the dominant gain. As evident from previously defined unit cell, consisting of two passive beam pairs and one active beam (2:1) (~80% expansion, **Figure *5(d)***, considering the maximum value). Within the lattice, midpoint-midpoint (denoted by $d_1$, $d_2$, $d_3$…,**Figure 7(b)**), corresponds to one beam pair and one active beam (1:1), providing a reduced 37% expansion. Therefore, if we instead consider the entire lattice, we have a total of five rows of active beams and six rows of rotating beams, resulting in a bulk expansion of 54% (which lies between 37% and 80%). Interestingly, it can be projected that the top and bottom passive beams provide an extra boost in the expansion. If these beams and half of the associated active beams are removed, we essentially approach the midpoint-to-midpoint value. As the number of unit cell rows increases, this boundary contribution converges toward zero. These results confirm that the multicellular structure effectively translates localized actuation into coordinated, large-scale deformation, validating the scalability of the design strategy.

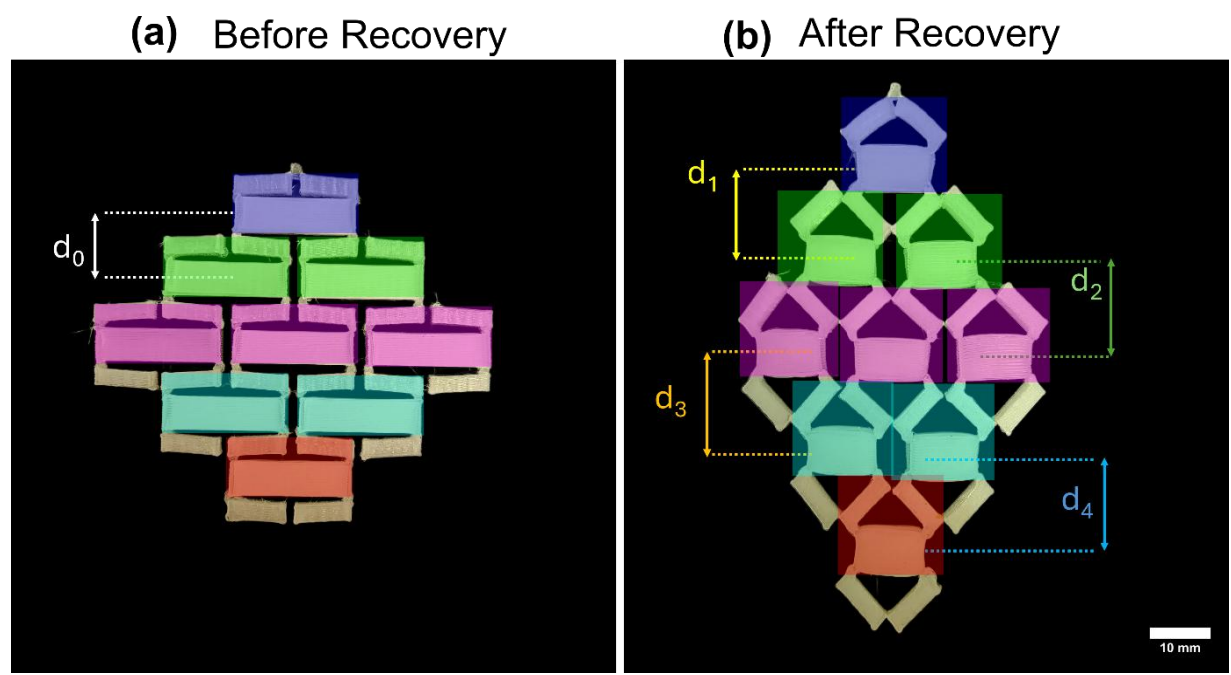

**Figure 7**

(a) and (b) Before and after recovery images of the lattice configuration. The color shaded regions represent unit cell(s) in each row forming a lattice. Midpoint-midpoint distance has been shown as well before and after recovery.

## 3. Conclusion

This study demonstrates the capabilities of the Programming via Printing (PvP) approach for creating architected structures with spatially varied trapped strain using a single functional material. Through systematic tuning of printing parameters and geometric design, we achieved multi-dimensional, programmable expansion in both unit cell and multicellular configurations. Experimental observations were in good agreement with geometrical modeling. Notably, the use of single-material simplifies fabrication and reduces cost, broadening the applicability of PvP in additive manufacturing of shape-shifting applications. Potentially, other additive manufacturing methods with higher resolution fabrication or high-volume fabrication platforms might be able to utilize our fabrication strategy and models for further levels of scaling or miniaturization. Hinge designs could be further optimized based on the specific printing strategy and application requirements. The current work did not address the change in mechanical responses after actuation of the active beam (or the orientation of the passive beams). Future work could explore mechanically tunable responses through more complex PvP strategies and design architectures, potentially enabling adaptive, reconfigurable systems for applications in soft robotics, deployable structures, and wearable devices.

## 4. Experimental Section

**SMP Filament Preparation and Thermal Characterization**

SMP filament was extruded from commercially available thermoplastic polyurethane (TPU) pellets (SMP MM3520, SMP Technologies Inc.) by a filament maker (Composer 450, 3Devo, Utrecht, The Netherlands) after drying in a vacuum oven at 50°C for 24 h. The filaments were stored in a vacuum-sealed storage to avoid moisture absorption. Thermal characterization by DSC was reported previously (Q200, TA Instruments, New Castle, DE, USA) [19].

**Fabrication of SMP samples for uniform strain trapping**

To print structures, STL files prepared in CAD were loaded in a commercially available slicer. The printing conditions were then programmed in the slicer. The files were then loaded into a conventional 3D printer with a 0.2 mm diameter nozzle (Ender 3 S1 direct drive, Creality, Shenzhen, China). Structures were then printed according to the defined programmed path, nozzle temperature, and printing speed. The print bed was set to the printing room ambient temperature of 20°C and was covered by heat-resistant Kapton tape for better adhesion. Samples were then recovered in water at 70°C for 30 min, which was higher than its glass transition temperature ($T_g$=35°C) to ensure faster recovery with no residual stress. Humidifiers were also employed to control the humidity in the printing room.

**Fabrication of SMP samples with non-uniform strain trapping**

The custom g-code files for non-uniform strain trapping were prepared. The key printing parameters, such as extrusion value, printing speed, programmed path, and nozzle temperature, were manipulated or added to adopt non-uniform printing conditions. Sacrificial elements were added at temperature transition zones to eliminate manual intervention during printing and prevent stringing or blobbing artifacts. Among the key printing parameters, the printing speed was 30 mm/s, the nozzle temperature varied between 195°C and 240°C, and the layer height was 0.2 mm unless otherwise stated. All structures were fabricated using a 0.2 mm nozzle (Ender 3 S1 direct drive, Creality, Shenzhen, China). Samples were recovered using the same conditions described above.

**Sample Size and Statistics**

All the sample groups were fabricated and tested three times (N = 3). For the SMP narrow strips, only one sample out of 54 (six printing speeds for each nozzle temperature, providing 18 groups in total) was discarded for showing abnormal outcomes and replaced by another sample within 72 hours of fabrication. In order to understand the effect of printing speed on the trapped strain, a one-way ANOVA with Tukey post hoc tests ($p < 0.05$) was conducted with printing speed as the only factor to test whether trapped strain differed significantly across speeds.

**Dimensional Measurement**

All the images were visually inspected with ImageJ. The trapped strain was measured as:

$$Trapped\ Strain = \frac{L_f - L_i}{L_i} \times 100\%$$

where $L_f$ = Final length after recovery, $L_i$ = Initial length after fabrication

Similarly, expansion was measured as:

$$Percentage\ of\ Expansion = \frac{d - d_o}{d_o} \times 100\%$$

The angles were designed in CAD and slicer software and were later measured again in ImageJ to confirm there was no deviation from the print path. For dimensional measurements, all measurements were made three times, and the average was recorded.

**Conflict of Interest**

The authors declare no conflict of interest.

**Data Availability Statement**

The data that support the findings of this study are available from the corresponding author upon reasonable request.

# Supporting Information

**Single-material 4D-printed shape-morphing structures via spatially patterned strain trapping**

S M Asif Iqbal[1,2], Hang Zhang[3], Lin Yang[2,4], Aoyi Luo[2,4], Joseph D. Paulsen[1,2,5], James H. Henderson[2,6]

[1] Department of Physics, Syracuse University, Syracuse, NY, USA
[2] Syracuse BioInspired Institute, Syracuse University, Syracuse, NY, USA
[3] Digital Building Technologies, Institute of Technology in Architecture, ETH Zurich, Zurich, Switzerland
[4] Department of Mechanical and Aerospace Engineering, Syracuse University, Syracuse, NY, USA
[5] Department of Physics, St. Olaf College, Northfield, MN, USA
[6] Department of Biomedical and Chemical Engineering, Syracuse University, Syracuse, NY, USA

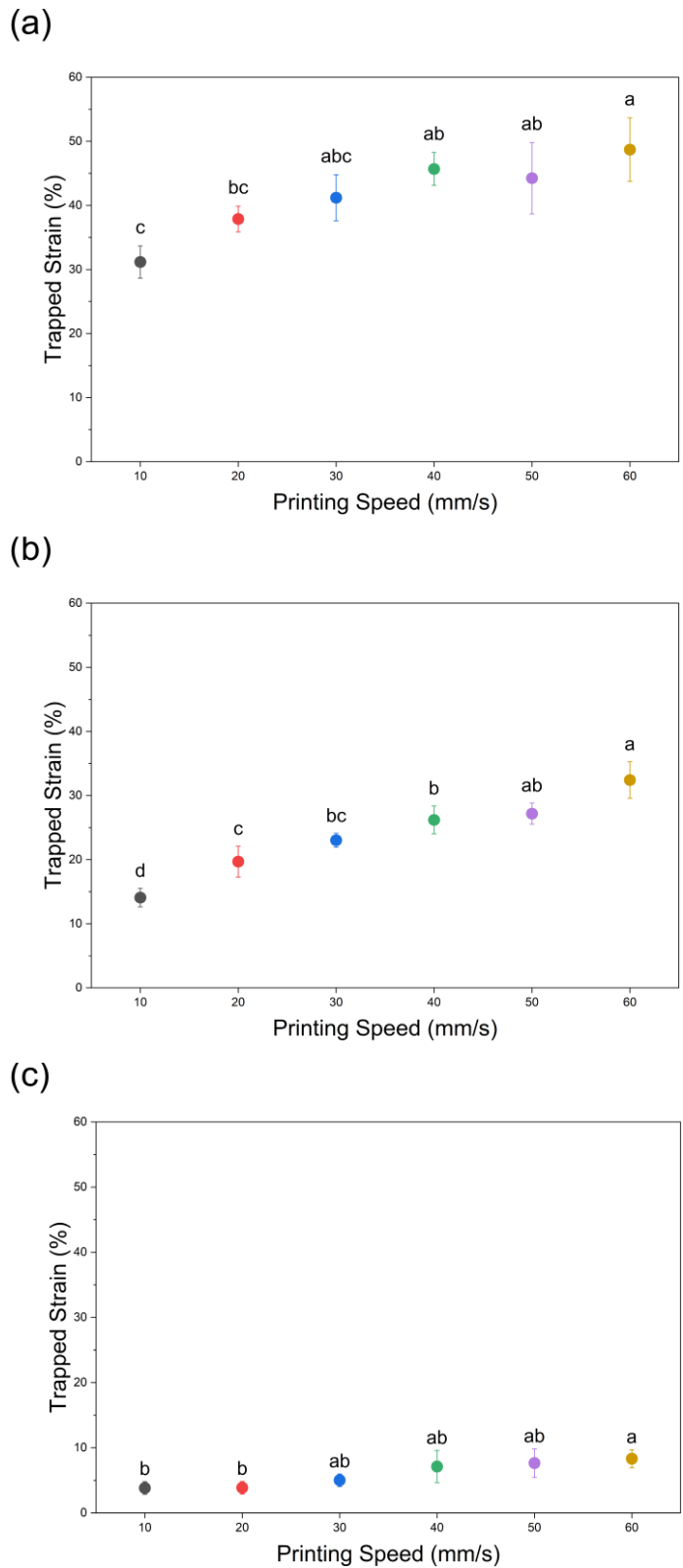

**Figure S 1**

Explicit plot of change in trapped strain with respect to print speed for different nozzle temperatures, (a) 195°C, (b) 215°C, and (c) 235°C, respectively. The letters represent statistically significant differences between the groups for different printing speeds. At 215°C, we have observed a clear increasing trend of trapped strain with printing speeds. For other cases, the trapped strain became steady after 30 mm/s.

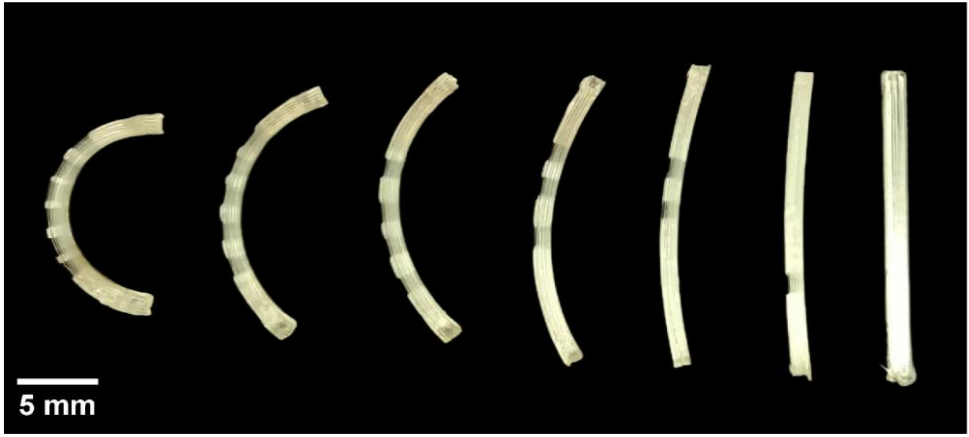

**Figure S 2**

The bending increased with the increase of print speed (from right to left). The very first sample on the right is the representative strip before recovery.

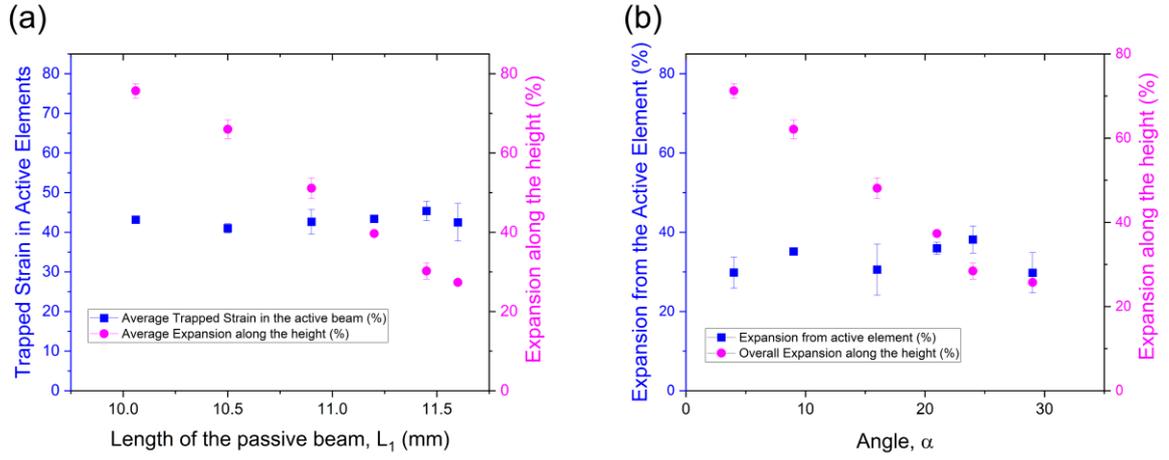

**Figure S 3**

(a) The trapped strain and (b) expansion in the active element for each case are shown explicitly for all six models.

**Modeling for the diamond-shaped unit cell**

As shown in **Figure 4(a)**, $L_0$ is the length of the center active element, and $L_1$ is the length of the outer passive element. The initial distance between the top and bottom points of the structure is $d_0$, and distance after actuation is $d$. Normalizing with respect to $L_0$, we define the following dimensionless parameters

$$L_1^* = \frac{L_1}{L_0}, \quad d_0^* = \frac{d_0}{L_0}, \quad d^* = \frac{d}{L_0}. \tag{S1}$$

Based on geometric relationships shown in **Figure 4(a), we have**

$$d_0^* = \sqrt{4(L_1^*)^2 - 1}, \tag{S2}$$

$$d^* = \sqrt{4(L_1^*)^2 - (1-\varepsilon)^2},  \tag{S3}$$

Let us consider

$$\alpha = d^* - d_0^*,  \tag{S4}$$

Differentiating **Equation (S4)** with respect to $\varepsilon$ yields

$$\frac{\partial \alpha}{\partial \varepsilon} = \frac{1-\varepsilon}{d^*} > 0.  \tag{S5}$$

Similarly, differentiate **Equation (S4)** with respect to $L_1^*$ yields

$$\frac{\partial \alpha}{\partial L_1^*} = \frac{4L_1^*}{d^*} - \frac{4L_1^*}{d_0^*} < 0.  \tag{S6}$$

Similarly, let us consider

$$\beta = \frac{d^*}{d_0^*} - 1,  \tag{S7}$$

Differentiating **Equation (S7)** with respect to $\varepsilon$ yields

$$\frac{\partial \beta}{\partial \varepsilon} = \frac{1-\varepsilon}{d_0^* d^*} > 0,  \tag{S8}$$

Similarly, differentiate **Equation (S7)** with respect to $L_1^*$ yields

$$\frac{\partial \beta}{\partial L_1^*} = \frac{4L_1^* \varepsilon(\varepsilon - 2)}{(d_0^*)^3 d^*} < 0.  \tag{S9}$$

Next, the beam width was taken into consideration. The geometric definition is shown in **Figure 4(b)**. Let us consider $H_0$ and $H_1$ represent the widths of the central active beam and the surrounding passive beams, respectively. The initial vertical distance $d_0$ is now given by the original distance predicted by the simplified model (neglecting beam width), denoted as $d_{0\_ori}$, plus an additional term accounting for the geometric contribution of beam width. Similarly, $d = d_{ori}$ (distance predicted by the model, neglecting the beam width) plus the additional terms from Poisson's effect and the geometric influence of beam width. Normalizing with respect to $L_0$, we define the following dimensionless parameters

$$H_0^* = \frac{H_0}{L_0}, \quad H_1^* = \frac{H_1}{L_0}, \quad L_1^* = \frac{L_1}{L_0}, \quad d_0^* = \frac{d_0}{L_0}, \quad d^* = \frac{d}{L_0},$$

$$d_{0\_ori}^* = \frac{d_{0\_ori}}{L_0} = \sqrt{4(L_1^*)^2 - 1}, \quad d_{ori}^* = \frac{d_{ori}}{L_0} = \sqrt{4(L_1^*)^2 - (1-\varepsilon)^2}.  \tag{S10}$$

Based on the geometric relationships shown in **Figure 4(b)** and the definitions of the dimensionless parameters in **Equation (S10)**, we obtain

$$d_0^* = d_{0\_ori}^* + H_0^* + \frac{H_1^*}{2L_1^*},  \tag{S11}$$

$$d^* = d_{ori}^* + H_0^*(1-\varepsilon)^{-\nu} + \frac{H_1^*(1-\varepsilon)}{2L_1^*}, \tag{S12}$$

where $\nu$ is Poisson's ratio.

Let us consider

$$\alpha = d^* - d_0^* \tag{S13}$$

Differentiating **Equation (S13)** with respect to $H_0^*$ yields

$$\frac{\partial \alpha}{\partial H_0^*} = (1-\varepsilon)^{-\nu} - 1 > 0. \tag{S14}$$

Similarly, differentiate **Equation (S13)** with respect to $H_1^*$ yields

$$\frac{\partial \alpha}{\partial H_1^*} = -\frac{\varepsilon}{2L_1^*} < 0. \tag{S15}$$

Differentiating **Equation (S13)** with respect to $L_1^*$ yields

$$\frac{\partial \alpha}{\partial L_1^*} = \frac{4L_1^*}{d_{ori}^*} - \frac{4L_1^*}{d_{0\_ori}^*} + \frac{H_1^*\varepsilon}{2(L_1^*)^2} \tag{S16}$$

Let us consider

$$\beta = \frac{d^*}{d_0^*} - 1. \tag{S17}$$

Differentiating **Equation (S17)** with respect to $H_0^*$ yields

$$\frac{\partial \beta}{\partial H_0^*} = \frac{d_0^*(1-\varepsilon)^{-\nu} - d^*}{(d_0^*)^2}. \tag{S18}$$

Similarly, differentiate **Equation (S17)** with respect to $H_1^*$ yields

$$\frac{\partial \beta}{\partial H_1^*} = \frac{d_0^*(1-\varepsilon) - d^*}{2L_1^*(d_0^*)^2} < 0. \tag{S19}$$

Differentiating **Equation (S17)** with respect to $L_1^*$ yields

$$\frac{\partial \beta}{\partial L_1^*} = \frac{d_0^*\left[\frac{4L_1^*}{d_{ori}^*} - \frac{H_1^*(1-\varepsilon)}{2(L_1^*)^2}\right] - d^*\left[\frac{4L_1^*}{d_{0\_ori}^*} - \frac{H_1^*}{2(L_1^*)^2}\right]}{(d_0^*)^2} \tag{S20}$$

**Modeling for the trapezoid unit cell:**

The trapezoidal unit cell consists of two trapezoids, in contrast to the diamond unit cell composed of two triangles. The geometric parameters are defined in **Figure S 4**, where only the bottom base of the trapezoid is the active element. The vertical distances between the top and bottom edges before and after recovery are denoted by $d_0$ and d, respectively. Normalizing with respect to $L_0$, we define the following dimensionless parameters

$$a^* = \frac{a}{L_0}, \quad L_1^* = \frac{L_1}{L_0}, \quad d_0^* = \frac{d_0}{L_0}, \quad d^* = \frac{d}{L_0}. \tag{S21}$$

Based on the geometric relationships shown in **Figure S 4** and the definitions of the dimensionless parameters in **Equation (S21)**, we obtain

$$d_0^* = \sqrt{(L_1^*)^2 - \frac{(1-a^*)^2}{2}}, \tag{S22}$$

$$d^* = \sqrt{(L_1^*)^2 - \frac{(1-\varepsilon-a^*)^2}{2}}. \tag{S23}$$

Let us consider

$$\alpha = d^* - d_0^*, \tag{S24}$$

Differentiating **Equation (S24)** with respect to $a^*$ yields

$$\frac{\partial \alpha}{\partial a^*} = \frac{1-\varepsilon-a^*}{2d^*} - \frac{1-a^*}{2d_0^*} < 0. \tag{S25}$$

For all admissible geometric ranges, the derivative is negative. Specifically: If a*<(1-ε)<1 both terms are positive, but the first term is smaller than the second, giving ∂α/∂a*<0. If (1-ε)<a*<1, the first term is negative and the second positive, yielding ∂α/∂a*<0. If (1-ε)<1<a, both terms are negative, and the magnitude of the first term is smaller than that of the second, again giving ∂α/∂a*<0. If a*=(1-ε)<1, the first term vanishes while the second remains positive, so ∂α/∂a*<0. If (1-ε)<1=a*, the second term vanishes while the first is negative, giving ∂α/∂a*<0. Therefore, ∂α/∂a*<0 holds for all cases, indicating that the recovery displacement decreases monotonically with increasing a*.

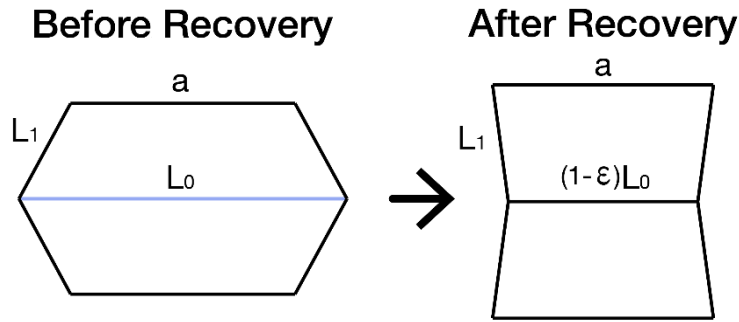

**Figure S 4**

The trapezoid design with side lengths $L_1$, top base of length a, and bottom base length $L_0$ before and after recovery.

## Modeling for the auxetic structure containing the unit cell

Consider the modeling including beam thickness, the geometric definition is shown in **Figure *6*(c)**. The width of the auxetic structure is *w*. Normalizing with respect to $L_2$, we define the following dimensionless parameters

$$w^* = \frac{w}{L_2}, \quad L_3^* = \frac{L_3}{L_2}, \quad d_0^* = \frac{d_0}{L_2}, \quad d^* = \frac{d}{L_2}, \quad H_3^* = \frac{H_3}{L_2}. \tag{S26}$$

where $H_3$ is the width of the outer element with length $L_3$. Based on the geometric relationships shown in **Figure *6*(c)** and the definition of the dimensionless parameters in **Equation (S26)**, we obtain

$$w^* = 2\left[\sqrt{1-(L_3^* - d^*)^2} + H_3^*\right] \tag{S27}$$

The relative change in the width before and after actuation is

$$\Delta w^* = 2\left[\sqrt{1-(L_3^* - d^*)^2} - \sqrt{1-(L_3^* - d_0^*)^2}\right], \tag{S28}$$

which suggests Δw is independent of $H_3$.

## Geometric parameters involved in diamond-shaped unit cell models

| Model | Angle, α | $L_0/2$ (mm) | $L_1/2$ (mm) |
|---|---|---|---|
| M1 | 4 | 10 | 10.06 |
| M2 | 9 | 10 | 10.5 |
| M3 | 14 | 10 | 10.9 |
| M4 | 19 | 10 | 11.2 |
| M5 | 24 | 10 | 11.45 |
| M6 | 29 | 10 | 11.60 |

*Table 1: Parameters involved in different unit cell models*

## Video S1 and Video S2

video/animation file (available upon request)